\documentclass[journal,10pt]{IEEEtran}
\usepackage{amsmath}
\usepackage{bm}
\usepackage{color,soul}
\usepackage{algorithmic}
\usepackage{hyperref}
\usepackage{threeparttable}
\usepackage{subfigure}
\usepackage{graphicx,epsfig,balance}
\usepackage{epstopdf}
\usepackage{cite}
\usepackage{color}
\usepackage{multirow}
\usepackage{amsfonts,amssymb}
\usepackage{booktabs}
\usepackage[table,xcdraw]{xcolor}
\usepackage[ruled]{algorithm2e}

\SetKwInOut{Input}{\textbf{Input}}
\SetKwInOut{Output}{\textbf{Output}}

\begin{document}
\title{Ultra Lite Convolutional Neural Network for Fast Automatic Modulation Classification in Low-Resource Scenarios}
\author{
Lantu Guo,
Yu Wang, \emph{Student Member}, \emph{IEEE},
Yun Lin, \emph{Member}, \emph{IEEE},\\
Haitao Zhao, \emph{Senior Member}, \emph{IEEE},  and
Guan Gui, \emph{Senior Member}, \emph{IEEE}

\IEEEcompsocitemizethanks{
\IEEEcompsocthanksitem Lantu Guo is with the School of Information and Electronics, Beijing Institute of Technology, Beijing 100811, China. He is also with the China Research Institute of Radiowave Propagation, Qingdao 266107, China (e-mail: guolantu@163.com). Yu Wang, Haitao Zhao, Guan Gui are with the College of Telecommunications and Information Engineering, Nanjing University of Posts and Telecommunications, Nanjing 210003, China (e-mail: 1018010407@njupt.edu.cn, zhaoht@njupt.edu.cn, guiguan@njupt.edu.cn).	Yun Lin is with the College of Information and Communication Engineering, Harbin Engineering University, Harbin 150000, China (e-mail: linyun@hrbeu.edu.cn).
}}

\markboth{IEEE Communications Letters, vol. xx, no. xx, pp. xx-xx, month, Year}{}
\maketitle

\begin{abstract}
Automatic modulation classification (AMC) is a key technique for designing non-cooperative communication systems, and deep learning (DL) is applied effectively to AMC for improving classification accuracy. However, most of the DL-based AMC methods have a large number of parameters and high computational complexity, and they cannot be directly applied to low-resource scenarios with limited computing power and storage space. In this letter, we propose a fast AMC method with lightweight and low-complexity using ultra lite convolutional neural network (ULCNN) consisting of data augmentation, complex-valued convolution, separable convolution, channel attention, and channel shuffle. Simulation results demonstrate that our proposed ULCNN-based AMC method achieves an average accuracy of 62.47\% on RML2016.10a and only 9,751 parameters. Moreover, ULCNN is verified on a typical edge device (Raspberry Pi), where the interference time per sample is about 0.775 ms. The reproducible code can be downloaded from GitHub\footnote{https://github.com/BeechburgPieStar/Ultra-Lite-Convolutional-Neural-Network-for-Automatic-Modulation-Classification}.
\end{abstract}


\IEEEpeerreviewmaketitle

\section{Introduction}
Automatic modulation classification (AMC) is a promising technology for analyzing the characteristics of signals, and it has wide applications in both military and civilian scenarios \cite{Dobre2007}. In recent years, deep learning (DL) has been introduced into AMC for automatically extracting efficient features from huge amount of samples by convolutional neural network (CNN) \cite{CNN}, recurrent neural network \cite{RNN}, graph neural network \cite{GNN} and so on, and its performance has been greatly improved.


T. O'Shea \emph{et al.} \cite{RML2016} firstly exposed a AMC dataset, i.e., RML2016.10a, and proposed a VT-CNN2 model for AMC. The effectiveness of DL for AMC is verified for the first time. After that, the goal of DL-based AMC methods is to design models with higher classification performance, faster computation speed, smaller model size, and fewer labeled samples for training \cite{Huangaccess2019,Accelerate2,HuangIoT2020,HuangIoT2020a,QiTCCN2020}. For better classification performance, F. Zhang \emph{et al.} \cite{MCNet} proposed a robust structure based on multiple convolutional blocks with residual connection and asymmetric convolution kernels, named as MCNet, which also achieved outstanding performance. J. Xu \emph{et al.} \cite{MCLDNN} proposed a hybrid structure with 1D convolution, 2D convolution and long short-term memory (LSTM), named as MCLDNN, and it can extract and fuse spatiotemporal features from the in-phase (I) component, quadrature (Q) component and both IQ component. It can outperform the previously proposed DL-based AMC methods. 

Based on MCLDNN, paper \cite{PET} proposed a lightweight and low-complexity structure, combining parameter estimation and transformation (PET), 2D convolutional layer, and gated recurrent unit (GRU), which is denoted as PET-CGDNN. It can have the similar performance with MCLDNN, but the model size of the former is only less than 1/5 of that of the latter.
X. Fu \emph{et al.} \cite{SCNN} introduced separable convolution to modify DL models, i.e., separable CNN (SCNN), for reducing parameters and computational complexity. However, simple modification can only reduce a certain amount of computation, and the performance of these models can not reach that of MCLDNN.

In this letter, we focus on the model designing of lightweight and low-complexity structure for low storage and computing resource scenarios. Therefore, a ultra lite CNN (ULCNN) model is proposed for realizing fast AMC. This model has the similar classification performance with MCLDNN, and its average accuracy can reach up to 62.47\% in RML2016.10a; It has equivalent computation speed to SCNN and MCNet, and the interference time per sample is less than 1 ms on a typical edge device, and almost 0.01 ms on GPU. More importantly, it has far smaller mode size than all existing DL models for AMC, which has only 9,751 parameters.

\section{Signal Model and Problem Formulation}
The equivalent complex baseband signal model is given as
\begin{equation}
\label{eq: CP}
\begin{split}
x(k) = h(k)*s(k) + n(k), k\in[1, K],
\end{split}
\end{equation}
where $x(k)$ represents the received signal, $h(k)$ represents the wireless channel, $s(k)$ represents the transmitted signal, and $n(k)$ represents the additive white Gaussian noise.
Hence, a complex baseband signal sample can be written as ${\bf x} = [x(1), x(2),\cdots, x(K)]$, and the DL-based AMC problem can be described as
\begin{equation}
\label{eq: AMC}
\begin{split}
\hat {\rm y} = \mathop{\arg\max}_{{\rm y} \in \bf Y}f({\rm y}|{\bf x}; {\bf W}),
\end{split}
\end{equation}
where ${\rm y}$ and $\hat {\rm y}$ are the ground-truth modulation type and the predicted modulation type, respectively; $\bf Y$ represents the modulation type pooling; $f({\bf W})$ is the mapping function from samples to modulation types, where ${\bf W}$ is the model weight. In the DL-based AMC method, the core is to design an efficient DL model (i.e., $f({\bf W})$) with high accuracy and low complexity. Thus, we propose a ULCNN model for AMC in the next section, and aim to achieve a lite structure with high accuracy, small model size and low computational complexity.

\section{The Proposed ULCNN-based AMC Method}
\begin{figure*}[htbp]
	\centering
	\includegraphics[width=7 in] {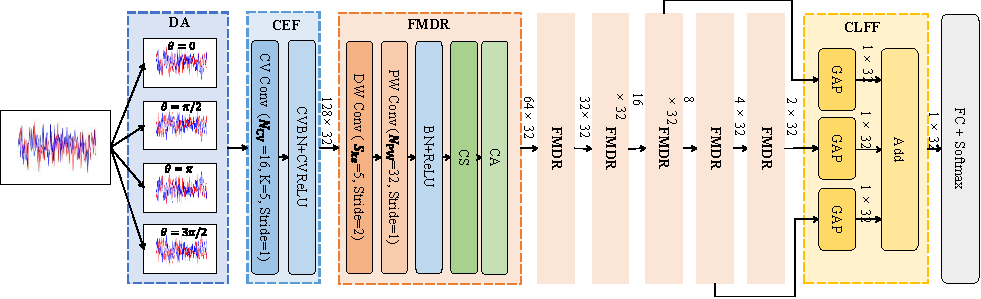}
	\caption{The overview of ULCNN for fast AMC method.}
	\label{fig:Structure}
\end{figure*}

The proposed ULCNN-based AMC method is shown in Fig. \ref{fig:Structure}, which consists of three key components, i.e., complex-valued (CV) convolution-based IQ channel fusion (IQCF) module, separable convolution-based feature mining and dimensionality reduction (FMDR) module, and cross-layer feature fusion (CLFF) module.
Firstly, the IQCF module, based on CV convolution, is applied to mine the correlation of in-phase and quadrature channels. Then, six cascaded FMDR modules, following behind the IQCF module, are applied for both extracting in-depth features and reducing the dimensionality of features. Finally, the CLFF module is used to fuse the different levels of in-depth features, and these fused features are fed into the last fully-connection (FC) layer for modulation classification.

In addition, there are two tricks for better performances, i.e., data augmentation (DA) and adaptive learning rate. The former is applied to expand the number of training samples in the pre-processing phase, and the latter is applied for automatically adjust the learning rate according to the training effect in the training phase.

\subsection{DA Method for Training Samples}
In this paper, rotation \cite{rotate1}, as a simple but efficient DA method, is used for augmenting samples in the training phase. The augmented sample can be written as
\begin{equation}
\label{eq: rotation}
\left[
\begin{array}{c}
\mathcal{R}({\bf \widetilde x})\\
\mathcal{I}({\bf \widetilde x})
\end{array}
\right] =
\left[
\begin{array}{cc}
{\cos} \theta&{-\sin} \theta\\
{\sin} \theta&{\cos} \theta
\end{array}
\right]
\left[
\begin{array}{c}
\mathcal{R}({\bf x})\\
\mathcal{I}({\bf x})
\end{array}
\right],
\end{equation}
where ${\bf \widetilde x}$ is the augmented training sample; $\theta\in\{0, \pi/2, \pi, 3\pi/2\}$ is the angle of rotation; $\mathcal{R}(\cdot)$ and $\mathcal{I}(\cdot)$ are the operations of obtaining the real and imaginary parts, respectively, and these two parts are also corresponding to the in-phase and quadrature components. Based on the DA, The number of the augmented samples for training is four time that of the original samples.

\subsection{The Key Components of ULCNN}
\subsubsection{IQCF module}
The IQCF module is based on CV convolution, and it aims to fuse the in-phase and quadrature components of signal samples and find the correlation between them \cite{CVN}. The formula of CV convolution can be written as
\begin{equation}
\label{eq: IQCF}
\begin{split}
{\bf W}_{\rm CV}*{\bf \widetilde x}&  = \mathcal{R}({\bf W}_{\rm CV})*\mathcal{R}({\bf \widetilde x}) - \mathcal{I}({\bf W}_{\rm CV})*\mathcal{I}({\bf \widetilde x})\\
&~+ j\cdot\left[\mathcal{R}({\bf W}_{\rm CV})*\mathcal{I}({\bf \widetilde x}) + \mathcal{I}({\bf W}_{\rm CV})*\mathcal{R}({\bf \widetilde x})\right],
\end{split}
\end{equation}
where ${\bf W}_{\rm CV}$ is the weight of CV convolution.
Assuming that there are $N_{\rm CV}$ neurons, the kernel size is $S_{ke}$, the convolution stride is 1, and the dimensionality of the input complex baseband signal samples is $K \times 1$, the number of the multiply-accumulate (MACC) operations about one CV convolution layer, which a typical measurement of the computation complexity, can be expressed as
\begin{equation}
\label{eq: MACC}
\begin{split}
N^{\rm MACC}_{\rm CV} = 4 \times K \times 1 \times N_{\rm CV} \times S_{ke}.
\end{split}
\end{equation}

There are CV batch normalization (CVBN) and CV rectified linear unit (CVReLU), following behind the CV convolution layer. Thus, the output feature map of the IQCF module is ${\bf M}_{\rm IQCF} = f_{\rm CVReLU}\left[f_{\rm CVBN}({\bf W}_{\rm CV}*{\bf \widetilde x}_{tr})\right]$, the dimensionality of that is $K \times N_{\rm CV}$.
In addition, considering that the following modules are based on real-valued (RV) operations, ${\bf M}_{\rm IQCF}$ needs to be converted into a RV matrix, and the converted output is $[\mathcal{R}({\bf M}_{\rm IQCF}), \mathcal{I}({\bf M}_{\rm IQCF})]^{\mathrm T}$ with dimensionality $K \times 2N_{\rm CV}$. Thus, under the condition of $N_{\rm CV}=16$ and $K=128$, the output dimensionality is  $128\times32$, rather than $128\times16$.

\subsubsection{FMDR module}
The main goal of the FMDR module is to extract the in-depth features and gradually reduce feature dimensionality. Here, the designed FMDR module consists of separable convolution, channel shuffle (CS) and channel attention (CA).
The separable convolution is a lightweight convolution with fewer parameters and lower computation complexity, when comparing with the standard convolution, and it consists of the depth-wise (DW) convolution and the point-wise (PW) convolution.
Assuming that the input of one FMDR module is ${\bf I}_{\rm FMDR}$ with dimensionality $N_{\rm FMDR} \times C_{\rm FMDR}$, the output feature map of the DW convolution can be expressed as
\begin{equation}
\label{eq: DW}
\begin{split}
{\bf M}_{\rm DW} = {\bf I}_{\rm FMDR}(:,c)*{\bf W}_{\rm DW}(:,c),~ c\in[1, C_{\rm FMDR}],
\end{split}
\end{equation}
where ${\bf W}_{\rm DW}$ is the weight of the DW convolution. Here, the convolution stride is set as 2, and the dimensionality can be reduced to ${N_{\rm FMDR}}/{2} \times C_{\rm FMDR}$. Moreover, the PW convolution is a standard convolution with the kernel size of 1 and the convolution stride of 1. The output feature map of the PW convolution is ${\bf M}_{\rm PW}={\bf M}_{\rm DW}*{\bf W}_{\rm PW}$, and is dimensionality is ${N_{\rm FMDR}}/{2} \times N_{\rm PW}$, where $N_{\rm PW}$ is the number of neurons in the PW convolution. It is noted that we set $N_{\rm PW} = 2N_{\rm CV}$, which is to ensure that the output channels of each layer are consistent for minimizing memory access cost \cite{Shufflenet}.

Here, the separable convolution is applied to reduce the computational complexity and the parameters. The number of the MACC operations about one separable convolution layer can be written as
\begin{equation}
\label{eq: MACC}
\begin{split}
N^{\rm MACC}_{\rm Sep} = &\frac{N_{\rm FMDR}}{2} \times C_{\rm FMDR} \times S_{ke}\\
&+ \frac{N_{\rm FMDR}}{2} \times C_{\rm FMDR} \times N_{\rm PW} \times 1.
\end{split}
\end{equation}

It is obvious that the computation complexity of the separable convolution layer is only ${1}/{N_{\rm PW}}+{1}/{S_{ke}}$ of that of the standard convolution layer (The number of neurons is $N_{\rm PW}$, the kernel size is $S_{ke}$, and the convolution stride is 2).
Following behind the separable convolution layer, CS and CA are applied for enhancing the classification performance. CS is an operation of disrupting the channel order of the input feature map, i.e., ${\bf M}_{\rm CS} = f_{\rm CS}({\bf M}_{\rm PW})$.

The principle of CA \cite{CAM} is given as follows. First, the feature map ${\bf M}_{\rm CS}$ is fed into both the global average pooling operation and the the global max pooling operation for two feature vectors; Then, these two feature vectors pass through a shared deep neural network (DNN) for their corresponding attention vectors, and this DNN has two FC layers with $\frac{N_{\rm PW}}{16}$ and $N_{\rm PW}$ neurons, respectively; Next, these two attention vectors are added and Sigmoid is applied to the superimposed vector for the final channel attention vector ${\bf V}_{\rm CA} = f_{\rm Sigmoid}\{f_{\rm DNN}[f_{\rm GAP}({\bf M}_{\rm CS})] + f_{\rm DNN}[f_{\rm GMP}({\bf M}_{\rm CS})]\}$; Finally, the output feature map can be written as
\begin{equation}
\label{eq: FMDR}
\begin{split}
{\bf M}_{\rm FMDR}(:, n) = {\bf M}_{\rm CS}(:, n)\cdot{\bf V}_{\rm CA}(n), n\in[1, N_{\rm PW}].
\end{split}
\end{equation}

\subsubsection{CLFF module}
CLFF module is applied to fuse these in-depth features from the fourth, the fifth and the sixth FDMR modules by add operation, which can be written as
\begin{equation}
\label{eq: ADD}
\begin{split}
{\bf M}_{\rm CLFF} = \sum_{l=4}^{6}f_{\rm GAP}({\bf M}_{\rm FMDR}^{l}),
\end{split}
\end{equation}
where ${\bf M}_{\rm CA}^{l}$ is the output feature map of the $l$-th FMDR module, and ${\bf M}_{\rm CLFF}$ is fed into a FC layer with Softmax activation function for classification.

\subsection{Loss Function and Optimizer with Adaptive Learning Rate}
Here, the loss function is cross entropy loss function, which is a common loss function for multi-category classification problem, and Adam with the adaptive learning rate (ALR) is applied as optimizer. The rule of adaptive learning rate is that if the validation loss cannot be reduced within ten epoches, the learning rate will decay to 80\% of the original.




\section{Simulation Results}
Here, the simulation environment is GTX1080Ti with Python3.7, Tensorflow1.14 and Keras2.1.4, and the simulation results are based on RML2016.10a. The dataset for this paper is RML2016.10a, where the number of the training, validation, and test samples is 77,000, 33,000, and 110,000, respectively. Moreover, $N_{\rm CV}$ is 16, while $N_{\rm PW}$ is 32; $S_{ke}$ is set as 5. In addition, the initial learning rate, the number of epoches and the batch size are 0.001, 200, and 128.

\subsection{Classification Performance and Convergence}
\begin{figure}[htbp]
  \centering
  \subfigure[Without DA]{
   \label{fig:acc1}
   \includegraphics[width=3.25 in] {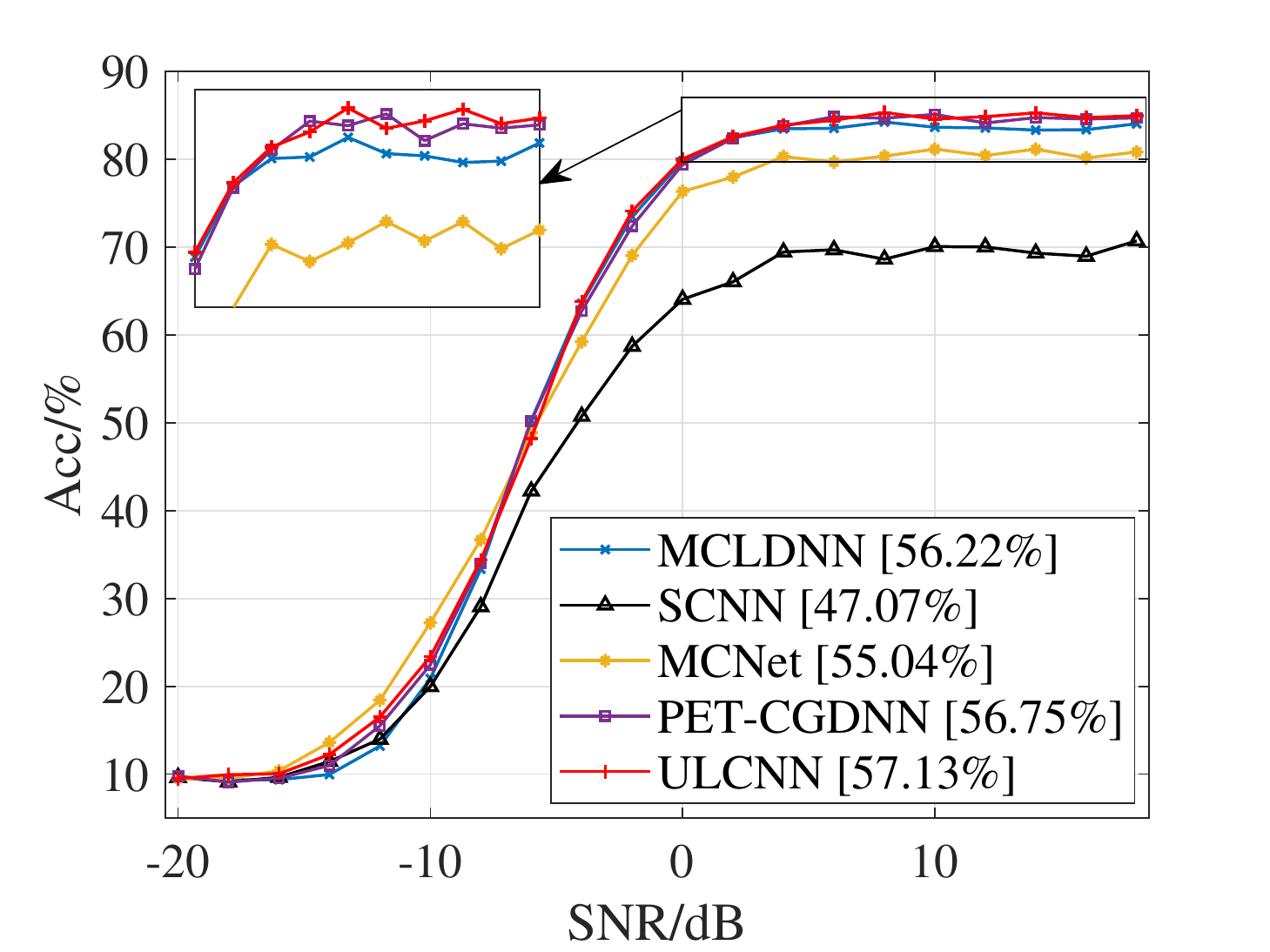}}
  \subfigure[With DA]{
   \label{fig:acc2}
   \includegraphics[width=3.25 in] {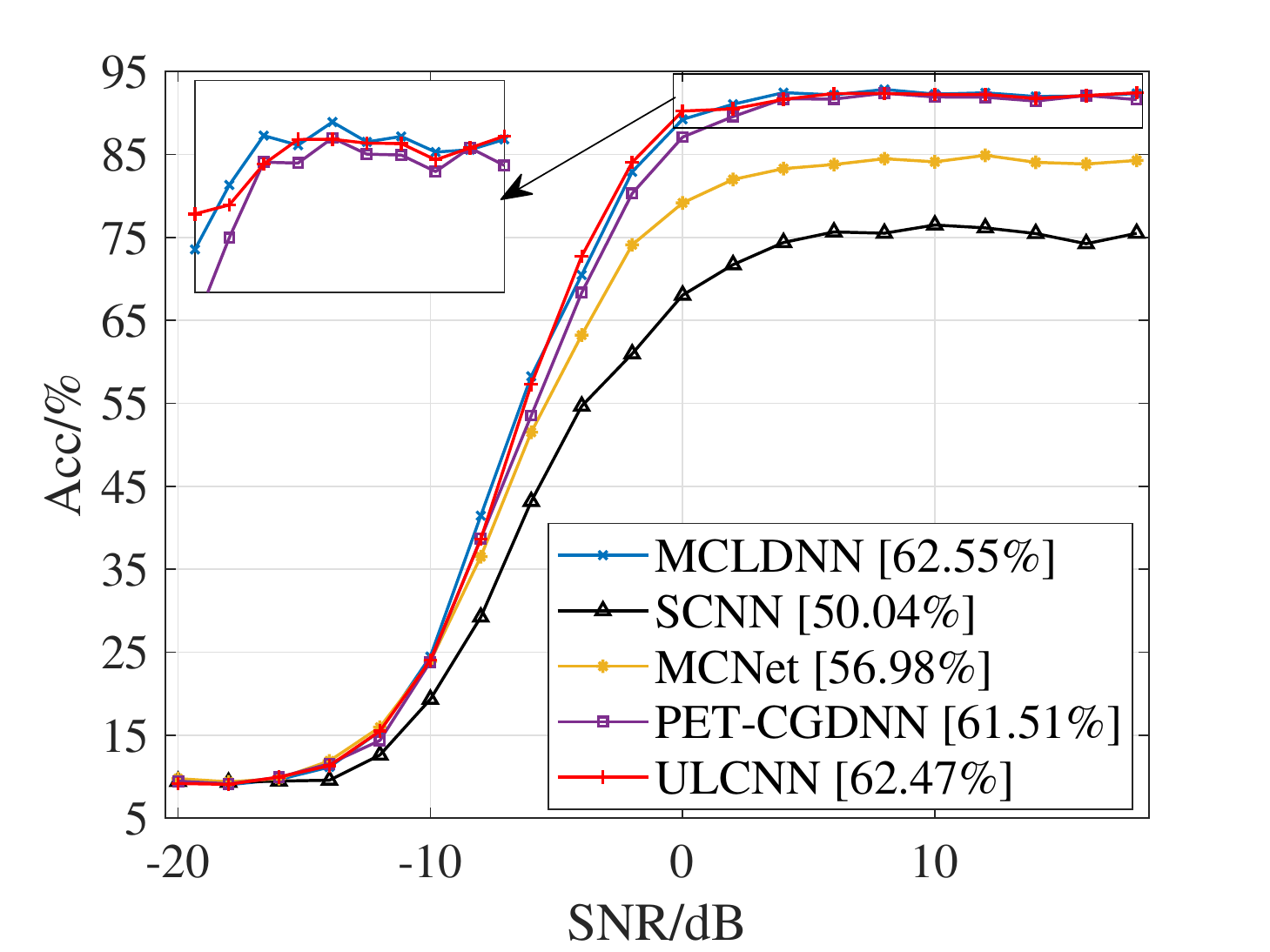}}
  \caption{The classification performance of different methods. It is noted that ``[**.**\%]'' represents the average classification performance.}
  \label{Acc}
\end{figure}

Here, the classification performance is given in Fig. \ref{Acc}. It is obvious that without the assistance of DA, our proposed ULCNN outperforms other comparison methods. In detail, the average accuracy of our proposed ULCNN at all SNRs can reach 57.13\%, and it has 0.91\% and 0.39\% improvement in the average classification performance, when comparing with MCLDNN and PET-GCDNN. Of course, the proposed ULCNN also outperforms MCNet and far exceeds SCNN in the classification performance.

More importantly, if DA is applied, these above five methods will achieve great performance improvement, and their improved average classification performance can reach 1.94\%$\sim$6.33\%, which demonstrates the importance of the number of training samples. In detail, MCLDNN has better performance than other methods, whose average classification accuracy can be 62.55\%, but our proposed ULCNN also achieves advanced performance, which only has 0.08\% average classification performance gap with MCLDNN. In addition, MCLDNN and ULCNN can both achieve high than 92\% classification accuracy at high SNRs (for instance, SNR $\geq$ 6dB), which is far higher than the classification accuracy of about 84\% and 85\% under the condition of no DA operation.

\begin{figure}[htbp]
	\centering
	\subfigure[Validation loss vs. epoch]{
		\label{fig:acc1}
		\includegraphics[width=3.25 in] {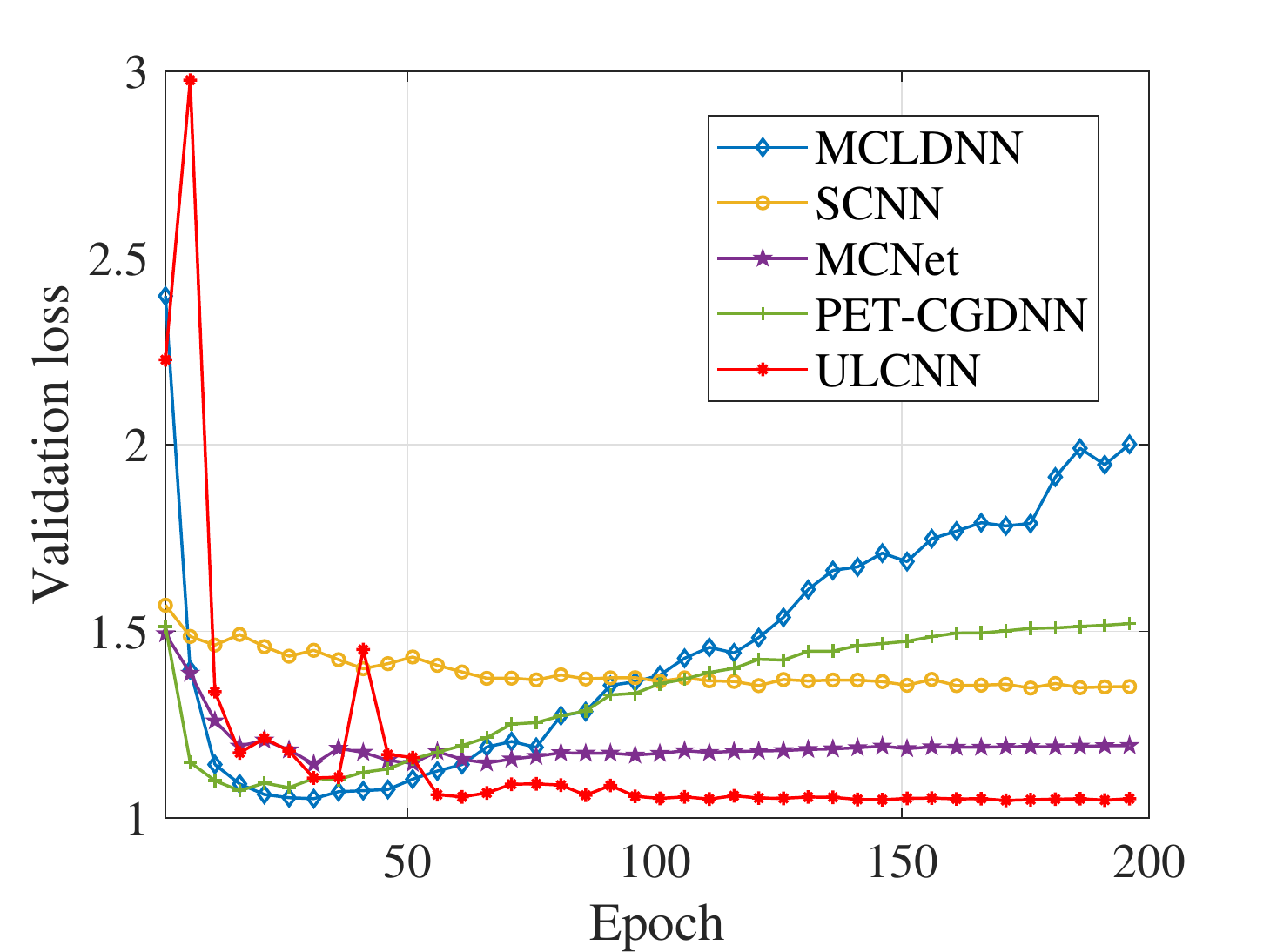}}
	\subfigure[Validation accuracy vs. epoch]{
		\label{fig:acc2}
		\includegraphics[width=3.25 in] {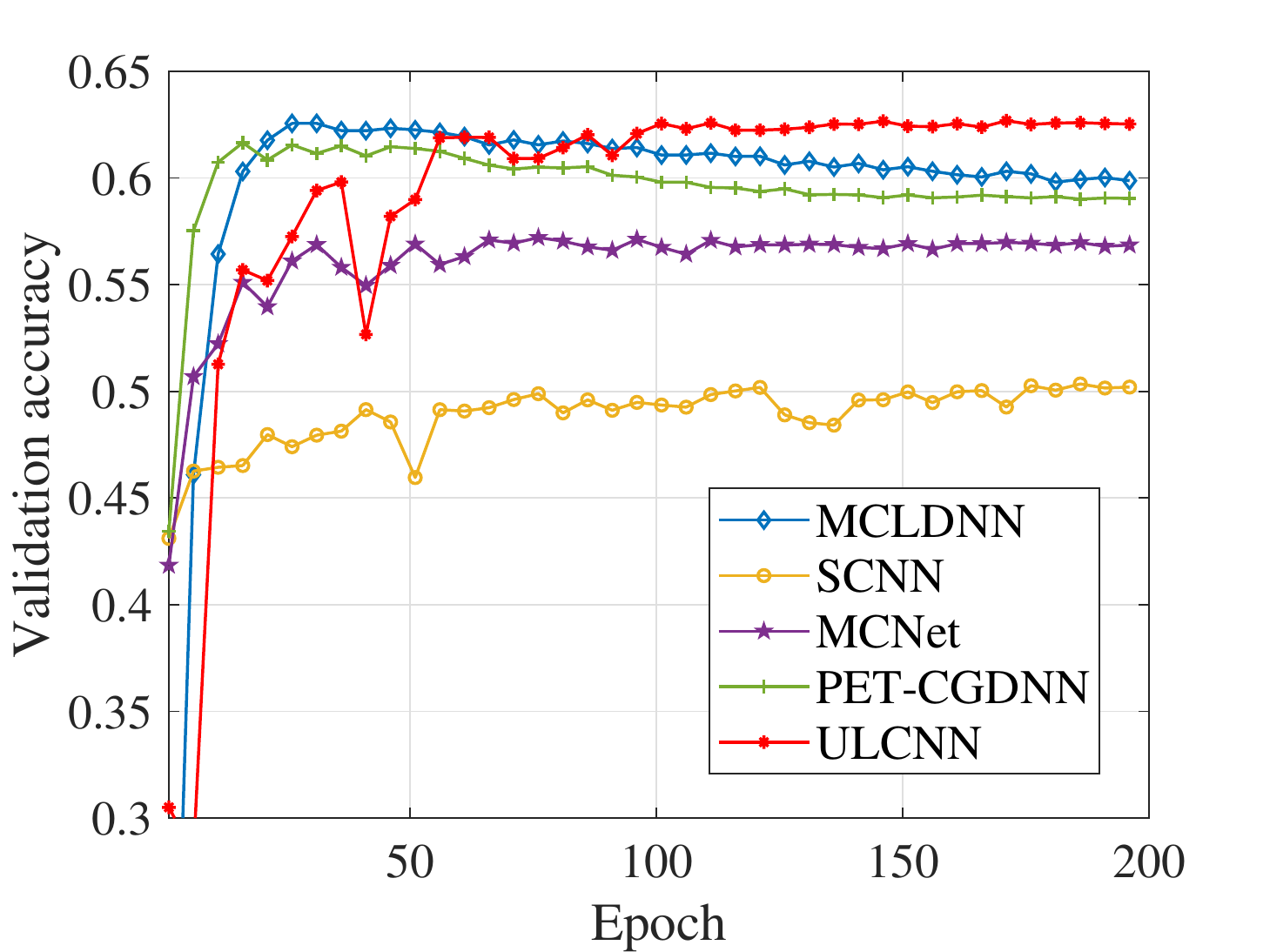}}
	\caption{The convergence of different models for AMC with DA.}
	\label{Loss}
\end{figure}

In addition, the convergence of different models is given in Fig. \ref{Loss}. It can be observed that MCLDNN and PET-CGDNN can reach the minimum loss and the maximum accuracy around the 30th epoch. However, in the subsequent training phase, their validation losses showed an upward trend and their validation accuracies showed a downward trend. This indicates that MCLDNN and PET-CGDNN have overfitting problems, while other methods, including ULCNN, have been steadily decreasing and have no overfitting trend even after reaching the minimum point or the maximum point. They continue to oscillate up and down near the minimum or maximum values.

\begin{table}[]
	\begin{center}
		\caption{The ablation studies.}
		\begin{tabular}{@{}ccc@{}}
			\toprule
			\multicolumn{1}{l}{Method} & \multicolumn{1}{l}{Ablation part} & \multicolumn{1}{l}{Average accuracy} \\ \midrule
			ULCNN                      & -                                 & 62.47\%                              \\
			ULCNN-A                    & w/o CV Conv                       & 61.94\% (0.53\%$\downarrow$)                    \\
			ULCNN-B                    & w/o CA                            & 61.56\% (0.91\%$\downarrow$)                    \\
			ULCNN-C                    & w/o CS                            & 60.92\% (1.54\%$\downarrow$)                    \\
			ULCNN-D                    & w/o CLFF                          & 62.19\% (0.28\%$\downarrow$)                    \\
			ULCNN-E                    & w/o ALR                           & 61.79\% (0.67\%$\downarrow$)                    \\ \bottomrule
		\end{tabular}
	\end{center}
	\begin{tablenotes}
		\item[1] *``w/o CV Conv'' represents that RV convolution replaces CV convolution, and in order to ensure consistent dimensionality of the output feature map, the number of neurons in RV convolution has doubled.
	\end{tablenotes}
	\label{Tab:AE}
\end{table}

\subsection{Ablation Studies}
The ablation studies are given in Tab. \ref{Tab:AE}, and six key components are discussed. Obviously, DA is the most important component in these component, which has been analyzed in the above section. Except DA, the CS component and the CA component also play great roles in the performance improvement, and their improved accuracies can reach up to 0.91\% and 1.54\%.
In addition, CV convolution-based IQCF module, CLFF module, and adaptive learning rate also have a certain effect on performance improvement, and the improved accuracies are only 0.53\%, 0.67\% and 0.28\%. The above simulation results show that all components in ULCNN are useful.

\begin{table*}
\caption{The complexity of different methods.}
\setlength\tabcolsep{3pt}
\begin{center}
\begin{tabular}{c|cc|cccc|cccc}
\hline
Methods   & $N^{\rm Para}$ & \begin{tabular}[c]{@{}c@{}}$N^{\rm MACC}$\\ ($\times 10^{6}$)\end{tabular} & \multicolumn{4}{c|}{\begin{tabular}[c]{@{}c@{}}Inference time per sample on GPU (unit: second)\\ (Batch size = 1/10/100/1000)\end{tabular}} & \multicolumn{4}{c}{\begin{tabular}[c]{@{}c@{}}Inference time per sample on edge device (unit: second)\\ (Batch size = 1/10/100/1000)\end{tabular}} \\ \hline
MCLDNN   & 406,199    & 18.10                                                 & 6.68$\times 10^{-3}$                     & 7.06$\times 10^{-4}$                    & 1.50$\times 10^{-4}$                    & 3.13$\times 10^{-5}$                    & 6.78$\times 10^{-2}$                      & 1.74$\times 10^{-2}$                     & 9.99$\times 10^{-3}$                      & 7.97$\times 10^{-3}$                      \\
SCNN      & 104,395    & 1.79                                                  & \textbf {2.22$\times 10^{-3}$}                     & \textbf {2.40$\times 10^{-4}$}                    & \textbf {3.85$\times 10^{-5}$}                  & 1.01$\times 10^{-5}$                    &  \textbf {3.89$\times 10^{-3}$}                      & 1.21$\times 10^{-3}$                      & 1.09$\times 10^{-3}$                      & 1.20$\times 10^{-3}$                      \\
MCNet     & 90,763     & 3.11                                                  & 4.18$\times 10^{-3}$                     & 4.42$\times 10^{-4}$                    & 7.17$\times 10^{-5}$                    & \textbf {8.40$\times 10^{-6}$}                    & 4.74$\times 10^{-3}$                     & \textbf{1.20$\times 10^{-3}$}                     & \textbf{8.25$\times 10^{-4}$}                     & \textbf{7.64$\times 10^{-4}$}                      \\
PET-CGDNN & 71,487     & 1.30                                                  & 3.46$\times 10^{-2}$                     & 3.55$\times 10^{-3}$                    & 3.70$\times 10^{-4}$                    & 4.12$\times 10^{-5}$                    & 2.15$\times 10^{-2}$                      & 5.11$\times 10^{-3}$                      & 2.54$\times 10^{-3}$                      & 1.89$\times 10^{-3}$                      \\
ULCNN (proposed) & \textbf{9,751}      & \textbf{0.20}                                                  & 6.31$\times 10^{-3}$                     & 6.27$\times 10^{-4}$                    & 5.96$\times 10^{-5}$                    & 1.08$\times 10^{-5}$                    & 5.85$\times 10^{-3}$                      & 1.33$\times 10^{-3}$                      & 8.84$\times 10^{-4}$                      & 7.75$\times 10^{-4}$                     \\ \hline
\end{tabular}
\end{center}
  \begin{tablenotes}
  \item[1] *The GPU is NVIDIA GTX 1080Ti, and the edge device is Raspberry Pi 4B with 8G RAM, and the above test is based on Python3.7, Tensorflow1.14 and Keras2.1.4.
  \end{tablenotes}
\label{Tab:Complexity}
\end{table*}

\subsection{Complexity Analysis}
In this paper, we give two metrics, the number of the MACC operations $N^{\rm MACC}$ (indirect metric) and the inference time per sample (direct metric), to describe the computational complexity, while the number of the parameters $N^{\rm Para}$ represents the model size. In addition, it is noted that we only calculate $N^{\rm MACC}$ about the main components, including the convolution layers, FC layers and attention operations, but some minor components with very less calculation, such as activation function, pooling and BN, are not considered. The detailed results are given in Tab. \ref{Tab:Complexity}.

Simulation results show that our proposed ULCNN only has less than 10,000 parameters, which is almost 1/40 and 1/7 of the parameters of MCLDNN and PET-GCDNN, and it is one of the DL models with the fewest parameters for AMC in RML2016.10a. Similarly, the theoretical computational complexity $N^{\rm MACC}$ of our proposed ULCNN is far less than that of other methods.
Considering that $N^{\rm MACC}$ can not directly reflect the inference speed of the model, the inference time per sample as the direct metric is also given. Here, these methods are tested on the GPU and the edge device, which represents the scenario with abundant computing power and scenario with restricted computing power, and the former is a GTX 1080Ti, while the latter is a Raspberry Pi 4B. The detailed results are given in Tab. \ref{Tab:Complexity}.

Unfortunately, our proposed method is not the fastest model, though its theoretical computational complexity is very low. When tested on the GPU, SCNN and MCNet have outstanding performance. Specifically, SCNN is faster than other methods when tested on GPU or on the edge device with small bach size, but MCNet can exceed SCNN, when the batch size increases to a certain extent. In addition, our proposed ULCNN has some disadvantages in testing on the GPU, but it has the similar inference speeds with MCNet in testing on the edge device, which demonstrates that our proposed method is more suitable for the scenario with restricted computing power.

Here, the main reasons why the theoretical computational complexity is seriously inconsistent with the actual inference speed is that:
\begin{itemize}
\item In the theoretical computational complexity, the convolutional layer and the FC layer account for a huge proportion. Although they consume most of the inference time, other operations, including data I/O, CS operation and element-wise operations (such as activation function, add operation and so on), also occupy a considerable amount of the inference time, and the latter is difficult to reflect in the theoretical computational complexity \cite{Shufflenet}. For instance, the theoretical computational complexity of PET-CGDNN is less than that of SCNN and MCNet, but the former have slower inference speed than the latter two. One of reasons is that although the PET operation is a low-computational-complexity operation, it takes up a lot of inference time.

\item The degree of parallelism of the model will also affect the inference time \cite{Shufflenet}. For instance, although MCNet has higher theoretical computational complexity than SCNN, PET-CGDNN and ULCNN, it is a highly parallel structure, and is faster than other methods.
\end{itemize}

\section{Conclusion}
In this paper, we propose a fast ULCNN-based AMC method by integrating various tricks to design a lightweight and low-complexity DL model. Simulation results show the performance superiority of our proposed AMC method, and it is very suitable for edge scenarios with limited computing power and storage space. In the future works, we aim to design hardware-friendly DL models to accelerate its interference by neural network pruning, neural network quantization, knowledge distillation, and so on.

\end{document}